\documentclass[aps,12pt,superscriptaddress,amsfonts,amssymb,amsmath]{revtex4-2}
\pdfoutput=1

\usepackage{graphicx}
\usepackage{epsfig}
\usepackage{makeidx}
\usepackage{epstopdf}
\usepackage{natbib}
\usepackage{xcolor}
\usepackage{amsmath}

\usepackage{booktabs}  
\usepackage{array}  
\usepackage{paralist}  
\usepackage{verbatim}
\usepackage{subfig}
\usepackage{braket}
\usepackage{appendix}
\usepackage{leftidx}
\usepackage{empheq}
\usepackage{dsfont}
\usepackage{hyperref}

\newcommand{\angstrom}{\mbox{\normalfont\AA}}
\newcommand{\xipzero}{{\xi'}}

\newcommand{\sumkr}{\sum_{p=1}^2\int d \Omega_{\hat k}}
\newcommand{\vecn}{n}
 
\newcommand{\loadingratio}{x} 

\begin{document}

\title{Coherent Plasma in a Lattice}

\author{L.\ Gamberale \footnote{Email address: luca.gamberale@gmail.com}}
\affiliation{Quantumatter Inc., Dover DE USA\\ LEDA srl, Università Milano Bicocca \\ I-20126 Milano, Italy}

\author{G.\ Modanese \footnote{Email address: giovanni.modanese@unibz.it}}
\affiliation{Free University of Bozen-Bolzano \\ Faculty of Science and Technology \\ I-39100 Bolzano, Italy}

\linespread{0.9}

\begin{abstract}
We present a fully second-quantized calculation showing the emergence of spontaneous coherent configurations of the electromagnetic field in interaction with charged bosons in a regular lattice. The bosons tend to oscillate at their plasma frequency, but are also subjected to electrostatic forces which keep them confined close to lattice sites and  cause a frequency shift in the oscillation. Under certain conditions on these frequencies, we find that a suitably defined set of coherent states (coherent both in the field and matter degrees of freedom) exhibit a negative energy gap with respect to the perturbative ground state. This is true in the RWA approximation and for position-independent fields, both to first and second order in the interaction Hamiltonian. We compare this result with other recent findings from cavity QED and notice that:
(1) consideration of full 3D wavefunctions and a careful definition of the coherent states are essential for obtaining the energy gap; 
(2) although our calculation is referred to bosons, it may also apply to protons bound in a crystal matrix, if their density is very low compared to the density of available states.

\end{abstract}

\maketitle

\section{Introduction}
The stability of the perturbative ground state in single-mode cavity QED has been intensively studied under wide conditions for quantum gases and electron systems \cite{rokaj2018light,andolina2019cavity,andolina2020theory,ashida2020quantum,guerci2020superradiant,stokes2020uniqueness,mivehvar2021cavity,roman2021photon,rokaj2022free}, either taking into account the diamagnetic term proportional to $\textbf{A}^2$ or not (the so-called ``no-go theorem'' for the superradiant transition, in its various versions, see \cite{andolina2019cavity, nogo2}).

There are also other important cases in quantum field theory in which the perturbative ground state is unstable, e.g.\ in QCD \cite{lauscher2000rotation,branchina1999antiferromagnetic} and in quantum gravity \cite{modanese1998stability,bonanno2013modulated,bonanno2019structure,modanese2021quantum}. Such instabilities are usually  related to a complex dynamical behavior not reducible to single-particle excitations.

In this work we consider an idealized physical system of localized oscillating charges coupled to a multi-mode monochromatic electromagnetic field, in which a finite energy gap develops at zero temperature, and we compute it rigorously using a set of trial states which are coherent both in the matter and field sectors. In the calculation the $\textbf{A}^2$ term is included and no-go theorems are eluded since we consider electromagnetic modes with momenta quantized in all possible directions. These modes are confined in the material due to a dispersion relation $k(\omega)$ that prevents them from escaping from it even in the absence of an external cavity.

There are of course some conditions that need to be satisfied for a negative gap, especially concerning the relation between the plasma frequency $\omega_p$ and another characteristic frequency $ \omega$ that defines the strength of the electrostatic potential bounding the charges to the lattice sites.

In Sect.\ \ref{Hamiltonian} we introduce the Hamiltonian of the system, which contains the kinetic term of the bosons in the presence of a vector potential plus the Hamiltonian of free cavity photons of momentum $\vec k$ (with $|\vec k| = \omega$) and an harmonic potential due to the combined effect of plasma oscillations and of local electrostatic ``cages'' near each lattice site. In real systems, these cages typically correspond to tetrahedral holes or octahedral holes \cite{burns1993mineralogical}. In agreement with standard approaches for handling the diamagnetic term \cite{faisal1987theory,rokaj2022free} we define new photon operators with the same momentum $\vec k$ and a shifted or ``dressed'' frequency $\omega'=\sqrt{\omega^2+\omega_p^2}$, equal to the oscillation frequency of the matter field (Appendix \ref{app:renorm}). The interaction term is written employing the dipole approximation (i.e., with the vector potential constant in space).

In Sect.\ \ref{selection} we define, via a canonical transformation, photon operators which project along the three space directions, in preparation for the introduction of trial coherent states $|\Omega \rangle$ in which matter oscillations occur in one specific direction (Sect.\ \ref{par:effective Hamiltonian}). The full Hamiltonian, when evaluated on these trial states in the rotating wave approximation (RWA), reduces to an effective Hamiltonian whose minimum exhibits an energy gap per particle of the form $\delta E^{(1)}_{\Omega}=\omega' |\alpha|^2 \left( 1- \frac{2\pi}{3} \varepsilon^2 \right)$, where $\varepsilon=\omega_p/\omega'$, so that the perturbative vacuum becomes unstable at first order when $\varepsilon > \varepsilon_{\text{crit}}\simeq 0.69$.

The factor $|\alpha|$ in the gap formula controls the oscillation amplitude in the trial coherent state, which is physically limited by the lattice spacing (Sect.\ \ref{sec:lowerbound}). In Sects.\ \ref{sec:secordpert}, \ref{sec:2ndorder} we compute the contribution to the energy gap given by the second perturbative order in the interaction Hamiltonian, describing processes in which photons are emitted and re-absorbed at different lattice sites. The resulting expression for the gap up to second order is very similar, namely $\delta E^{(2)}_{\Omega}=\omega' |\alpha|^2 \left( 1- \frac{8\pi}{3}\varepsilon^2 \right)$, implying condensation 
for even lower values of the coupling constant
$\varepsilon > \varepsilon_{\text{crit}}\simeq 0.35$.

In Appendix \ref{estimate} we give a numerical estimate of the condensation threshold and of the energy gap for a realistic case, namely for protons absorbed into a FCC metallic matrix, when their density is so low that they can be described by a bosonic wavefunction. Supposing one proton per lattice site and a lattice spacing $d=2.5\angstrom$ we obtain a coupling constant $\varepsilon=0.56 > \varepsilon_{\text{crit}}\simeq 0.35$. Thus the system is over threshold and condensation occurs under robust conditions. Assuming that the amplitude of the oscillation is fixed by the size of the octahedral sites of the crystal, the average energy gap per particle is $\delta E^{(2)}_{\Omega}\simeq -1$ eV and the frequencies $\omega$ and $\omega_p$ are in the THz range, in good agreement with values of the largest phonon frequencies adapted to the proton mass \cite{wang2004thermodynamic}.

In Sect.\ \ref{par:SpatDim} we estimate the spatial dependence of the field and we argue that coherence domains will form, at whose boundary the field decreases as the spherical Bessel function $j_0(r)$.

Sect.\ \ref{discussion} contains a final discussion and outlook to future work. From the historical point of view, we point out that extensive pioneering work on e.m.\ coherence in condensed matter was done in the 1990's by G.\ Preparata \cite{ref5}, who also introduced the concept of coherence domains. Our present approach was inspired by Preparata's work but is independent from it and completely self-consistent.

The main results of the paper are summarized in Sect.\ \ref{conc}.

In addition to the mentioned Appendices \ref{app:renorm} and \ref{estimate}, Appendices \ref{app:expval} and \ref{app:definition} contain some technical details of the calculation. In Appendix \ref{app:jellium} a static charge distribution is defined which reproduces the almost-harmonic potential used in the Hamiltonian.

\section{Hamiltonian}
\label{Hamiltonian}
\begin{figure} 
\centering
\includegraphics[width=.4\textwidth]{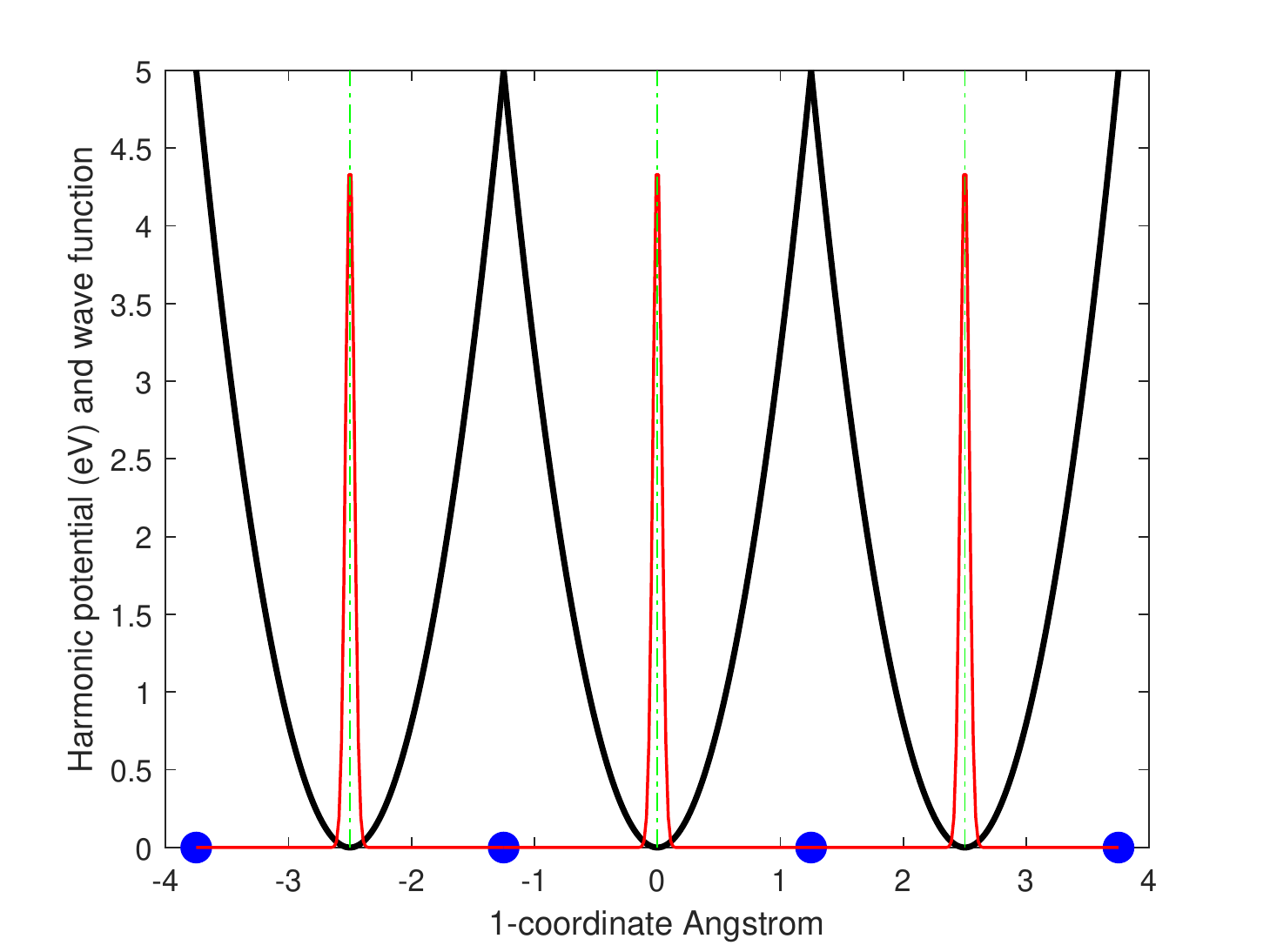}
\includegraphics[width=.4\textwidth]{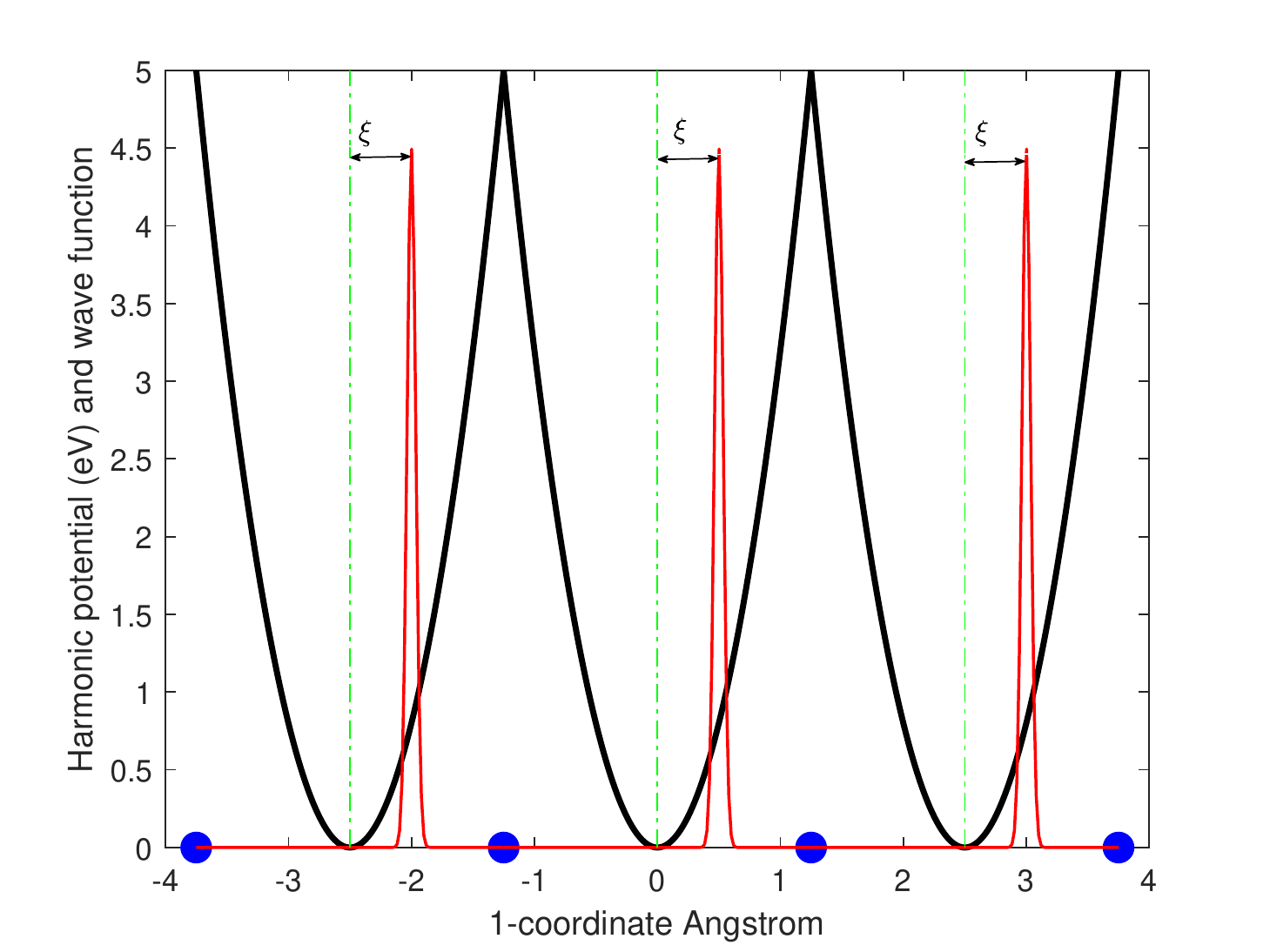}
\caption{Periodic harmonic potential in the 1-direction and square modulus of the wave functions  of the oscillators in their ground state (left) and elongated by $\xi$ due to their coherent oscillation (right).}
\label{fig:Harmonic}
\end{figure}
Let us consider a system of charged identical quantum oscillators placed at the vertices of a simple cubic three-dimensional lattice with spacing $d$. We suppose that the oscillator $\vecn$ is in its equilibrium position at site $\vecn$ of the  lattice, so that there are $N$ oscillators in a lattice with $N$ elementary cells. Oscillator $\vecn$ is kept in its equilibrium position by a  potential $V_{\vecn}(\vec\xi_{\vecn})$ which is harmonic with frequency $\omega$ when $|\vec\xi_{\vecn}|$ is small enough and becomes stiffer when the oscillation approaches the boundary of the elementary cell. This requirement guarantees that the oscillation will not interfere with the adjacent elementary cells and remains confined in the $\vecn$-th cell. 

The Hamiltonian of the single harmonic oscillator is (we will use natural units with $\hbar=c=k_B=\varepsilon_0=1$)
\begin{equation}
H^{(1)}_\text{osc}=\frac{[\vec p+e\vec A(\vec\xi,t)]^2}{2m}+\frac{m\omega^2}{2}\vec\xi^2
\label{eq:hclassical}
\end{equation}
where $e$, $m$ and $\omega$ are the charge, mass and oscillation frequency of the oscillator at position $\vec\xi$ and $\vec A(\vec\xi,t)$ is the electromagnetic vector potential.

Due to the self-generated electrostatic interaction the particles experience an additional potential given by $V_p(\vec\xi)=\frac{m\omega_p^2}2\vec\xi^2$, where $\omega_p=\sqrt{\frac{e^2N}{mV}}$ is the free plasma frequency.

The total Hamiltonian describing the oscillators in interaction with the electromagnetic field is therefore 
\begin{equation}
H^{(N)}_\text{osc}=\sum_{n=1}^{N}
\frac{[\vec p_n+e\vec A(\vec x_n+\vec\xi_n,t)]^2}{2m}+\frac{m(\omega^2+\omega^2_p)}{2}\vec\xi_n^2
\label{eq:hclassical1}
\end{equation}
where $\vec x_n$ is the equilibrium position of the $n$-th oscillator.

By adding to \eqref{eq:hclassical1} the free electromagnetic Hamiltonian, the total Hamiltonian can be rewritten in second-quantized terms as
\begin{equation}
\hat H_{\text{tot}}=
\hat H+
\hat H_{\text{int}}+
\hat H_{\text{photon}}
\label{eq:hsecquant0}
\end{equation}
where
\begin{subequations}
\label{eq:hsecquant1}
\begin{align}
\hat H=&
\omega'
\sum_{n=1}^N
\left[
\vec a_{\vecn}^\dagger(t) 
\cdot
\vec a_{\vecn} (t)
+\frac32\right]
\label{eq:hsecquanth0}
\\
\hat H_{\text{int}}=&
\sum_{n=1}^N
\frac em\hat{\vec p}_n\cdot\hat{\vec A}(\vec x_n+\vec\xi_n,t)
\label{eq:hsecquanth1}
\\
\hat H_{\text{photon}}=&
\omega
\sumkr
\left[
b_{p,\vec k}^\dagger(t) 
b_{p,\vec k} (t)
+\frac12\right]+
\sum_{n=1}^N
\frac{e^2}{2m}\hat{\vec A}^2(\vec x_n+\vec\xi_n,t)
\label{eq:hsecquanth2}
\end{align}
\end{subequations}
and where we have defined the shifted oscillation frequency 
\begin{equation}
    \omega'=\sqrt{\omega^2+\omega_p^2}.
\end{equation}
Note that for the electromagnetic field we have selected the modes with frequency $\omega$  only, since we are interested in the modes that have a relevant dynamical impact on the system under study in a way that will be clear in the following. The neglected modes interact with the system only perturbatively.

The destruction and creation operators $a_{\vecn}(t)$, $a_{\vecn}^\dagger(t)$, $b_{p,\vec k}(t)$ and $b_{p,\vec k}^\dagger(t)$ are in the interaction picture, so that
\begin{equation}
\left\{
\begin{aligned}
a_{\vecn} (t)=&a_{\vecn}e^{-i\omega't}
\\
a_{\vecn}^\dagger(t)=& a_{\vecn}^\dagger e^{i\omega't}
\\
b_{p,\vec k}(t)=&b_{p,\vec k}e^{-i\omega t}
\\
b_{p,\vec k}^\dagger(t)=&b_{p,\vec k}^\dagger e^{i\omega t}
\end{aligned}
\right.
\label{eq:intrep}
\end{equation}
and the radiation field operator is given by
\begin{equation}
\hat{\vec A}(\vec x,t)=
\frac{1}{\sqrt {2\omega V}}
\sumkr
[b_{p,\vec k}(t)e^{i\vec k\cdot\vec x}\vec\varepsilon_{p,\vec k}+b_{p,\vec k}^\dagger(t)e^{-i\vec k\cdot\vec x}\vec\varepsilon^*_{p,\vec k}]
\label{eq:radiation}
\end{equation}
where $|\vec k|=\omega$, $\vec\varepsilon_{p,\vec k}\cdot\vec k=0$ and $\vec\varepsilon^*_{p,\vec k}\cdot\vec\varepsilon_{p',\hat k}=\delta_{pp'}$.
Here $p$ is a polarization index and the e.m.\ wave unit vector $\hat k=\vec k/\omega$ spans all possible 3D directions.
In order to get rid of the diamagnetic term  $\frac{e^2}{2m}\sum_{n=1}^N\hat{\vec A}^2(\vec\xi_n,t)
$ we will proceed in the context of the dipole approximation. Following the lines described in Appendix \ref{app:renorm} \cite{rokaj2022free}, we can recast Eq. \eqref{eq:hsecquanth2} in terms of new photon operators so that
\begin{equation}
\hat H_{\text{photon}}=
\omega'
\sumkr
\left(c^\dagger_{p,\vec k}c_{p,\vec k}+\frac12\right).
\label{eq:Hprenorm}
\end{equation}
Eq. \eqref{eq:Hprenorm} shows that the transformation \eqref{eq:cantrans} makes the diamagnetic term disappear, making the electromagnetic Hamiltonian diagonal. The oscillation frequency is shifted to the new value $\omega'$, equal to that of the matter field.
 Moreover, the electromagnetic modes acquire a mass and their dispersion relation gets modified so that the electromagnetic field is not able to propagate through the vacuum and cannot escape from the material.

The last term of the Hamiltonian can now be written in terms of the new operators. By means of the canonical transformations
\begin{equation}
\begin{aligned}
\vec a=&
\frac1{\sqrt2}
\left(
\frac 1{\xipzero}\vec \xi + i\xipzero\vec p
\right)
\\
\vec a^\dagger=&
\frac1{\sqrt2}
\left(
\frac 1{\xipzero} \vec\xi - i\xipzero\vec p
\right)
\end{aligned}
\label{eq:can}
\end{equation}
and
\begin{equation}
\begin{aligned}
\vec \xi=&\xipzero\frac{\vec a+\vec a^\dagger}{\sqrt2}
\\
\vec p=&\frac i\xipzero\frac{\vec a^\dagger-\vec a}{\sqrt2}
\end{aligned}
\label{eq:can1}
\end{equation}
where we have defined $\xipzero=(m\omega')^{-\frac12}$,
we can recast Eq. \eqref{eq:hsecquanth1} as
\begin{equation}
\hat H_{\text{int}}=
\frac em
\frac{1}{\sqrt {2\omega'V}}
\frac i{\xipzero}
\sum_{n=1}^N
\frac{\vec a_{\vecn}^\dagger(t)-\vec a_{\vecn}(t)}{\sqrt2}
\cdot
\sumkr
[c_{p,\vec k}(t)e^{i\vec k\cdot\vec x}\vec\varepsilon_{p,\vec k}+c_{p,\vec k}^\dagger(t)e^{-i\vec k\cdot\vec x}\vec\varepsilon^*_{p,\vec k}]
\label{eq:interaction}
\end{equation}
Since we are considering em modes whose wavelength is much larger than the lattice constant, we are allowed to neglect the spatial dependence of the electromagnetic field ({\it dipole approximation}, DA) so that Eq. \eqref{eq:interaction} becomes
\begin{equation}
\hat H_{\text{int}}=
\frac{i\omega_p}{2\sqrt {N}}
\sum_{n=1}^N
(\vec a_{\vecn}^\dagger(t)-\vec a_{\vecn}(t))
\cdot
\sumkr
[c_{p,\vec k}(t)\vec\varepsilon_{p,\vec k}+c_{p,\vec k}^\dagger(t)\vec\varepsilon^*_{p,\vec k}]
\label{eq:interaction1}
\end{equation}
where the coupling strength turns out to be proportional to the plasma frequency $\omega_p$.

In the next section we will go through a variational calculation of the minimum energy of the dynamical system described by Eqs. \eqref{eq:hsecquanth0},  \eqref{eq:interaction1},  and \eqref{eq:Hprenorm} when probed by matter-photon coherent sates.

\section{Selection of the quantum state for the electromagnetic field}
\label{selection}

The modes of the electromagnetic field that are eligible for resonant interaction with the oscillators are those whose frequency equals $\omega'$, precisely those that have been selected in Eq. \eqref{eq:interaction1}. The number of such independent modes is $2\cdot4\pi$, distributed along the possible directions of the electromagnetic momentum $\omega'\hat k$ and electric polarization.
We note here that we are now diverging from the usual treatment of the electromagnetic field in resonant cavities since our analysis is not limited to the wave vectors in a single direction but we consider the contribution of all the wave vectors whose modulus is $|\vec k| = \omega'$. This fact is bound to have dramatic consequences on the final result.

Assuming without loss of generality that the oscillation of the charges is along the direction $\hat 1$, we can define the photon states
\begin{equation}
\ket{\omega',\hat1}
=
\sqrt{\frac{3}{8\pi}}
\sumkr
\braket{\hat 1|\vec\varepsilon_{p,\vec k}}
\ket{\vec k,p}
\label{eq:photonstate}
\end{equation}
where the integral is performed over the directions of the unit vector $\hat k$.
These states are properly normalized thanks to the relation
\begin{equation}
    \sumkr |\braket{\hat 1|\vec\varepsilon_{p,\vec k}}
|^2=\int d \Omega_{\hat k}\bra{\hat 1}[\mathds{1}-\ket{\hat k}\bra{\hat k}]\ket{\hat1}=4\pi-\int d \Omega_{\hat k}\cos^2\theta_{\hat k}=\frac{8\pi}{3}.
\end{equation}
 
The states $\ket{\omega',\hat1}$ are created and annihilated by the operators
\begin{subequations}
 \begin{empheq}[left={\empheqlbrace\,}]{align}
C_1^\dagger
=&
\sqrt{\frac{3}{8\pi}}
\sumkr
\braket{\hat 1|\vec\varepsilon_{p,\vec k}}^*
c^\dagger_{\vec k,p}
\label{eq:Ba}
\\
C_1
=&
\sqrt{\frac{3}{8\pi}}
\sumkr
\braket{\hat 1|\vec\varepsilon_{p,\vec k}}
c_{\vec k,p}.
\label{eq:Bb}
\end{empheq}
\label{eq:B}
\end{subequations}
By introducing the operators
\begin{subequations}
 \begin{empheq}[left={\empheqlbrace\,}]{align}
C_2^\dagger
=&
\sqrt{\frac{3}{8\pi}}
\sumkr
\braket{\hat 2|\vec\varepsilon_{p,\vec k}}^*
c^\dagger_{\vec k,p}
\label{eq:Ca}
\\
C_2=&
\sqrt{\frac{3}{8\pi}}
\sumkr
\braket{\hat 2|\vec\varepsilon_{p,\vec k}}
c_{\vec k,p}
\label{eq:Cb}
\\
C_3^\dagger
=&
\sqrt{\frac{3}{8\pi}}
\sumkr
\braket{\hat 3|\vec\varepsilon_{p,\vec k}}^*
c^\dagger_{\vec k,p}
\label{eq:Da}
\\
C_3
=&
\sqrt{\frac{3}{8\pi}}
\sumkr
\braket{\hat 3|\vec\varepsilon_{p,\vec k}}
c_{\vec k,p}
\label{eq:Db}
\end{empheq}
\label{eq:CD}
\end{subequations}
the commutation relations for the new operators are
\begin{equation}
[C_i,C_j^\dagger]=
{\frac{3}{8\pi}}
\sumkr
\braket{\hat i|\vec\varepsilon_{p,\vec k}}
\braket{\vec\varepsilon_{p,\vec k}|\hat j}
=\delta_{ij},\ \ \ [C_i^\dagger,C_j^\dagger]=[C_i,C_j]=0 
\end{equation}
so that Eqs. \eqref{eq:B} and \eqref{eq:CD} are canonical transformations.
The vector potential after the DA can be rewritten in terms of these operators as
\begin{equation}
\hat{\vec A}=
\frac{1}{\sqrt {2\omega'V}}
\left[
(C_1+C_1^\dagger)\hat1+
(C_2+C_2^\dagger)\hat2+
(C_3+C_3^\dagger)\hat3
\right].
\label{eq:radiation2}
\end{equation}

We can now express the interaction and photon Hamiltonians in terms of the new operators. We get immediately from Eq. \eqref{eq:interaction1}
\begin{equation}
\hat H_{\text{int}}=
\frac {i\omega_p}{2\sqrt N}
\sqrt{\frac{8\pi}{3}}
\sum_{n=1}^N
[a_{1, \vecn}^\dagger C_1-a_{1, \vecn} C_1^\dagger
+a_{1, \vecn}^\dagger C_1^\dagger-a_{1, \vecn} C_1]
\label{eq:interaction2}
\end{equation}
and from Eq. \eqref{eq:Hprenorm}
\begin{equation}
\begin{split}
\hat H_{\text{photon}}=&
\omega'
\sum_{i=1}^3
\left[
C_i^\dagger C_i
+\frac12
\right]
\label{eq:EMconB}
\end{split}
\end{equation}

Eqs. \eqref{eq:interaction2} and \eqref{eq:EMconB} tell us that the different modes of the electromagnetic field contribute to increase the coupling with the matter field and such an interaction does not affect the energy density of the electromagnetic field. We will see in the next sections how this fact avoids the {\it no-go} theorem \cite{andolina2019cavity, nogo2} and leads to the instability of the perturbative vacuum and to the migration of the field  to a new stable, coherent configuration.

\section{Effective Hamiltonian \label{par:effective Hamiltonian}}
Let us consider the trial coherent state
\begin{equation}
\ket\Omega=
\bigotimes_{i=1...3;\vecn}
\ket{\alpha_i}_{\vecn}
\bigotimes_{j=1...3}
\ket{\mathcal A_j}
\label{eq:trial}
\end{equation}
with the definitions
\begin{subequations}
\label{eq:defstates}
\begin{align}
\ket{\alpha_i}_{\vecn}=&
e^{-\frac12|\alpha_i|^2}
\sum_{m=0}^\infty
\frac{\alpha_i^m}{m!}
a_{i,\vecn}^{\dagger m}\ket{0}_{\vecn}
\label{eq:defa}
\\
\ket{\mathcal A_j}=&
e^{-\frac12N|\mathcal A_j|^2}
\sum_{p=0}^\infty
\frac{(\sqrt N\mathcal A_j)^p}{p!}
C_j^{\dagger p}\ket{0}.
\label{eq:defB}
\end{align}
\end{subequations}
The state $\ket\Omega$ is a coherent state with a vector order parameter $\vec \alpha$ for the matter sector and $\sqrt N\vec{\mathcal A}$ for the electromagnetic sector. We make a further choice setting $\alpha_2=\alpha_3=\mathcal A_2=\mathcal A_3=0$ so that
\begin{subequations}
\label{eq:EA2}
\begin{align}
\vec a_{\vecn}\ket\Omega=&
\hat 1\alpha \ket\Omega
\label{eq:EA2a}
\\
\vec C\ket\Omega=&
\sqrt N\hat 1\mathcal A \ket\Omega
\label{eq:EA2b}
\end{align}
\label{eq:EA1}
\end{subequations}
The two order parameters $\alpha(t)$ and $\mathcal A(t)$ are complex functions that will be treated as variational parameters as well as the solution of the emerging dynamical equations.
The all important feature of the state $\ket\Omega$ that the parameter $\alpha$ is the same for all $\vecn$ allows us to drop the index $\vecn$ from the operators so that  Eqs. \eqref{eq:hsecquanth0},  \eqref{eq:interaction1},  and \eqref{eq:EMconB} can be rewritten as
\begin{subequations}
\label{eq:EA3}
\begin{align}
\bra\Omega H\ket\Omega=&
N\omega'\left[
|\alpha|^2 
+\frac32
\right]
\label{eq:EA3a}
\\
\bra\Omega H_{\text{int}}\ket\Omega=&
\frac {i\omega_pN}{2}
\sqrt{\frac{8\pi}{3}}
[\alpha^* \mathcal A-\alpha\mathcal A^* +
\alpha^* \mathcal A^*-\alpha\mathcal A ]
\label{eq:EA3b}
\\
\bra\Omega H_{\text{photon}}\ket\Omega=&
\omega'
\left[
N|\mathcal A|^2+\frac32\right].
\label{eq:EA3c}
\end{align}
\end{subequations}

Eqs. \eqref{eq:EA3} lead to the Schr\"odinger-like equation
\begin{equation}
    \bra\Omega i\frac{\partial}{\partial t}\ket\Omega=
{\bra\Omega H_{\text{tot}}\ket\Omega}.
\label{SL}
\end{equation}

By setting $\alpha(t)=|\alpha|e^{-i\Tilde\omega t}$, $\mathcal A(t)=|\mathcal A|e^{-i(\Tilde\omega t-\frac{\pi}{2}) }$ and $\varepsilon=\frac{\omega_p}{\omega'}$, Eqs. \eqref{eq:EA3} add up to
\begin{equation}
{\bra\Omega H_{\text{tot}}\ket\Omega}^{(1)}=
N\omega'
\left[ 
|\alpha|^2 
+\frac32
-
\sqrt{\frac{8\pi}{3}}
\varepsilon
|\alpha \mathcal A|+
|\mathcal A|^2+\frac3{2N}
\right]+\mathcal{H}_{crt}(t),
\label{eq:EA3d}
\end{equation}
where 
\begin{equation}
    \mathcal{H}_{crt}(t)=
    N\omega'
    \sqrt{\frac{2\pi}{3}}
\varepsilon
|\alpha \mathcal A|
(e^{2i\Tilde\omega t}+e^{-2i\Tilde\omega t})
\end{equation}
are the \emph{counter-rotating terms} that wildly oscillate at the double of the frequency $\Tilde{\omega}$ and where the superscript $^{(1)}$ indicates that the calculation is at first order in perturbation theory. 

We now introduce the \emph{rotating wave approximation} (RWA), consisting in neglecting the term $\mathcal{H}_{crt}(t)$ in the effective hamiltonian.
We note that the counter-rotating terms of the type $ab$ and $a^\dagger b^\dagger$, once evaluated on the trial state produce terms proportional to $A(t) \alpha(t)$ and $A^*(t) \alpha^*(t)$ which have a time dependence of the type $\exp (\pm 2 i\Tilde{\omega} t)$ and are averaged to zero over an oscillation period.
This does not happen for the terms retained, which have a temporal dependence of the type $A^*(t) \alpha(t)$ and $A(t) \alpha^*(t)$. They are time-independent and are therefore the only ones to make a contribution to the energy of the trial state. For this reason it is legitimate to apply the RWA approximation in our calculation, even though it is not generally valid in cavity QED in the presence of ultra-strong coupling \cite{frisk2019ultrastrong}.
 
Equation (\ref{SL}) then gives
\begin{equation}
\Tilde{\omega}=
N\omega'
\left[ 
|\alpha|^2 
+\frac32
-
\sqrt{\frac{8\pi}{3}}
\varepsilon
|\alpha \mathcal A|+
|\mathcal A|^2+\frac3{2N}
\right].
\label{eq:schroe}
\end{equation}

We are now equipped with a set of legitimate matter-field quantum states that satisfy the dynamical equations depending parametrically on $|\alpha|$ and $|\mathcal A|$.
Our goal is to look for solutions with energy content lower than that of the perturbative vacuum $|\alpha|=|\mathcal A|=0$. 
To this end we define the energy gap per particle as $\delta E^{(1)}_\Omega= \frac{1}{N}{\bra\Omega H_{\text{tot}}\ket\Omega}^{(1)}-\omega'
\left[ 
\frac32+\frac3{2N}
\right]$ and find the minimum with respect to the parameter $|\mathcal A|$ by requiring
\begin{equation}
\frac{\partial\delta E^{(1)}_\Omega}{\partial|\mathcal A|}=0.
\label{eq:min1}  
\end{equation} 
The minimum is found for $|\mathcal A|=\sqrt{\frac{2\pi}{3}}|\alpha|$ and the minimum of the energy gap is
\begin{equation}
\delta E^{(1)}_\Omega=
\omega'
|\alpha|^2
\left(1-\frac{2\pi}{3}
\varepsilon^2
\right).
\label{eq:EA3e}
\end{equation}

Eq. \eqref{eq:EA3e} shows that the vacuum becomes unstable when  
\begin{equation}
\varepsilon>\varepsilon_{\text{crit}}
=
\sqrt{\frac{3}{2\pi}}\simeq0.69
\label{eq:critcond}
\end{equation}
 which is well within the allowed range $0<\varepsilon<1$. 

This result reveals the existence of a phase transition to a new vacuum of the system. However, by looking at Eq. \eqref{eq:EA3e}, it appears that the new vacuum has a negatively diverging gap, due to the freedom of the variational parameter $|\alpha|$, clearly a non-physical situation.
Fortunately, due to the intervention of other factors not considered in the hamiltonian that limit the oscillations of the charges outside their reticular cages, the parameter $|\alpha|$ cannot assume arbitrarily large values making the result of Eq. \eqref{eq:EA3e} meaningful and rich of physical consequences. 

To be more specific, we must consider that the Hamiltonian \eqref{eq:hsecquanth0} is a quadratic approximation of the true Hamiltonian which is valid for small oscillations around the equilibrium positions of the oscillating charges which fails when the oscillation amplitude exceeds a certain value. The complete Hamiltonian contains terms that prevent the oscillations from exceeding the dimensions of the reticular cage. The detailed form of such terms is unknown, since it depends on the details of the charge distributions of the host lattice. We will introduce the effect of these terms by imposing a maximum value on the amplitude of oscillation of the charges which will be fixed by the size and shape of the lattice cells.
We will address these topics in the following paragraphs.

It is important to note that the symmetry breaking of the vacuum is made possible by the synergistic cooperation of the various modes of the electromagnetic field that makes the coupling to the matter field $\sqrt{\frac{8\pi}{3}}$ times that of the single mode (Eq. \eqref{eq:interaction2}). As far as we know, this feature is up to now a novelty in the literature dedicated to the study of the strong coupling regime.
Had we only considered the two polarization modes and the two directions $\hat k$ and $-\hat k$ along a fixed axis of the field, we would not have obtained a sufficiently intense coupling with the matter to induce the transition. In fact, the factor $\sqrt{\frac{8\pi}{3}}$ would have been substituted by $\sqrt{2\cdot2}$ (given by the two polarizations and the two directions) and Eq. \eqref{eq:EA3e} would have been substituted by the similar equation
\begin{equation}
\delta E^{(1)}_\Omega=
\omega'
|\alpha|^2
\left(
1-\varepsilon^2
\right)
\label{eq:EA3ewrong}
\end{equation}
that displays no instability for $\varepsilon$ in the allowed range $(0,1)$ and reproduces the result of various {\it no-go} theorems in the literature \cite{andolina2019cavity, nogo2}.
It is therefore imperative to consider the photon state \eqref{eq:photonstate} in order to achieve the symmetry breaking.


\section{Lower bound of the energy gap for the coherent phase \label{sec:lowerbound}}
When condition \eqref{eq:critcond} is met, Eq. \eqref{eq:EA3e} is unbounded from below because $|\alpha|$ can assume any value, an obviously  non-physical condition. Little thought is required to spot the direction we need to take to recover a physically sound result. The parameter $|\alpha|$ is related to the maximum amplitude of the plasma oscillation of the charges. It is clear that such an oscillation cannot take arbitrarily large values since it is limited by the linear size of the lattice ``cages'' whose spacing is $d$. Mathematically, this may be formulated by introducing an-harmonic terms into the matter Hamiltonian that are negligible for small plasma oscillations and become predominant when the oscillation exceeds the cage dimension. To implement this condition we need to fix a maximum value to the oscillation amplitude of the matter field, in general given by a fraction $f$ of the linear dimension $d$ of the electrostatic cages. Indeed by using Eq. \eqref{eq:can1} we must have
\begin{equation}
\frac{{\xipzero}^2}2
\max_t
\bra \Omega{(a_n(t)+a_n^\dagger(t))^2}\ket \Omega
=\frac{f^2d^2}4
\end{equation}
that, recalling Eq. \eqref{eq:EA2a}, becomes
\begin{equation}
|\alpha|^2
=\frac{f^2d^2}{8{\xipzero}^2}-\frac14
=\frac{mf^2d^2\omega'}{8}-\frac14
\label{eq:maxalpha}
\end{equation}

In all practical situations the last term in Eq. \eqref{eq:maxalpha}, stemming from the indetermination principle, can be neglected (being $d^2m \omega'>>2$) and the energy gap per particle gets the form 
\begin{equation}
\delta E_\Omega^{(1)}=
\frac{mf^2d^2{\omega'}^2}{8}
\left(
1-\frac{2\pi}{3}\varepsilon^2
\right).
\label{eq:firstorder}
\end{equation}

It is also important to note that had we used plane wave functions in place of coherent states we would not have been able to use any arguments relating to the non-divergence of the energy gap thus obliging to declare the solutions found as non-physical.

To summarize this section, we have found that when an ensemble of charged particles embedded in a neutralizing charge density of opposite charges reaches a critical density, it is subjected to a spontaneous quantum phase transition to a stable, collective and coherent state strongly and resonantly coupled  to a coherent electromagnetic field with a total energy lower than that of the incoherent state (uncorrelated particles and no macroscopic electromagnetic field) by a finite amount.

This configuration involves a very large number of particles and has a spatial extension much larger than the typical atomic radius. 
In Section \ref{par:SpatDim} we will make an estimate of the spatial structure and will introduce the key concept of coherence domain (CD).

\section{Second order Perturbation theory for composite coherent states \label{sec:secordpert}}
We now want to proceed with the analysis of the energy content of the trial state \eqref{eq:trial} by studying the second order contribution of the perturbative expansion of the ground state energy.
In physical terms, the first order contribution takes into account the contribution to the energy of the photons emitted by the oscillators and reabsorbed by the same oscillators, whereas the second order contribution considers the dispersive contribution of photons emitted by an oscillator and absorbed by a different oscillator. As will be shown below,
it turns out that the two terms contribute the same amount, thus doubling the
term of negative interaction and strengthening the condensation mechanism.

To apply the second order perturbation theory we need an orthonormal basis over which we can expand our trial state.
To this end, we use the property of the interaction term $H_{\text{int}}$
of preserving the sum of the number of photons and the excitation number of the
oscillators (the creation of a photon involves the reduction of the excitation state of the oscillator and vice versa) and define the new normalized basis as
\begin{equation}
\ket{\sigma, \alpha, \mathcal A}_{\vecn}
=
\frac{\sqrt{\sigma!}}{R^\sigma}
\sum_{\eta=0}^\sigma
 \frac{\alpha^{\sigma-\eta}}{\sqrt{(\sigma-\eta)!}}
 \frac{N^{\eta/2}\mathcal A^{\eta}}{\sqrt{\eta!}}
 \ket{\sigma-\eta}_{\vecn}\otimes\ket\eta
\end{equation}
where we have defined $R=
\sqrt{|\alpha|^2+N|\mathcal A|^2}
$
so that the trial state can be rewritten
\begin{equation}
\ket \Omega=
e^{-\frac12R^2}
\sum_{\sigma=0}^\infty
\frac{R^\sigma}{\sqrt{\sigma!}}
\bigotimes_{\vecn}
\ket{\sigma, \alpha, \mathcal A}_{\vecn}.
\label{eq:composite}
\end{equation}

The properties of the states $\ket{\sigma, \alpha, \mathcal A}_{\vecn}$ are the following:
\begin{itemize}
\item
${_{\vecn}}{\braket{\sigma, \alpha, \mathcal A|\sigma', \alpha, \mathcal A}}{_{\vecn'}}= \delta_{\sigma \sigma'}\delta_{nn'}$ 
\item
The expectation value of $H_{\text{int}}$ (Eq. \eqref{eq:interaction2}) on the states $\ket{\sigma, \alpha, \mathcal A}_{\vecn}$ is given by (see Appendix \ref{app:expval})
\begin{equation}
{_{\vecn}}{\bra{\sigma, \alpha, \mathcal A}}
H_{\text{int}}
{\ket{\sigma', \alpha, \mathcal A}}{_{\vecn'}}=
- 
\omega_p
\sqrt{\frac{8\pi}{3}}
|\alpha\mathcal A |
\frac{\sigma}{R^2}
\delta_{\sigma \sigma'}
\delta_{\vecn\vecn'}
\label{eq:Esigma1}
\end{equation}
\item
The expectation values of the unperturbed Hamiltonian is  (see Appendix \ref{app:expval})
\begin{equation}
\bra{\sigma, \alpha, \mathcal A}H
+\hat H_{\text{photon}}
\ket{\sigma, \alpha, \mathcal A}=
\omega'
\left[
N\left(\sigma\frac{|\alpha|^2}{R^2}
+\frac32+
|\mathcal A|^2
\frac{\sigma}{R^{2}}
\right)+\frac32
\right]
\label{eq:Esigma0}
\end{equation}

\end{itemize}

We now are ready to compute the second order contribution to the energy.

\section{The second order contribution \label{sec:2ndorder}}
We start from the well-known second order perturbation theory in the Brillouin-Wigner approximation \cite{BrillouinWigner}
 \begin{equation}
E_\sigma=
h_{0p}
+
h_1
+
\frac{h_1^2}
{E_\sigma-(
h_{0p}
+
h_1
)}
\label{eq:SOC1b}
\end{equation}
where we have indicated for simplicity $\ket {\sigma}=\bigotimes_{\vecn}
\ket{\sigma, \alpha, \mathcal A}_{\vecn}$, $h_{0p}=\bra {\sigma} H+\hat H_{\text{photon}}\ket {\sigma}$ and $h_1=\bra {\sigma} H_{\text{int}}\ket {\sigma}$.

To be strict, the numerator of Eq. \eqref{eq:SOC1b} should be $\frac{N(N-1)}{N^2}h_1^2$ since the sum over the unperturbed states should exclude the diagonal terms but, being $N$ very large, we are allowed to set $N(N-1)/N^2\simeq1$.

The solutions of Eq. \eqref{eq:SOC1b} are
\begin{equation}
E_\sigma=
h_{0p}
+
(1\pm1)h_1
\end{equation}
and the lowest energy per particle of the states $\ket \sigma$ is $E_\sigma=
h_{0p}
+
2h_1$. 
Substitution of Eqs. \eqref{eq:Esigma0} and \eqref{eq:Esigma1} into Eq. \eqref{eq:hsecquant0} yields at second order
\begin{equation}
\begin{split}
\frac{1}{N}
{\bra\Omega H_{\text{tot}}\ket\Omega}^{(2)}
&=
e^{-R^2}
\sum_{\sigma=0}^\infty
\frac{R^{2\sigma}}{\sigma!}
\frac{E_\sigma}N
=
\\
&=
\omega'
\left[
|\alpha|^2 
+\frac32+\frac3{2N}
-
\sqrt{\frac{8\pi}{3}}
2\varepsilon
|\alpha \mathcal A|+
|\mathcal A|^2
\right].
\end{split}
\label{eq:gap0}
\end{equation}
Eq. \eqref{eq:gap0} shows that at second order the contribution of the negative interaction term is twice the contribution at the first order so that the coupling threshold gets further lowered compared to the first order calculation. By repeating the procedure already performed in Sect.\ \ref{par:effective Hamiltonian} we find
\begin{subequations}
\begin{align}
|\mathcal A|=&\sqrt{\frac{8\pi}{3}}|\alpha|
\label{eq:Avsalpha}
\\
\delta E^{(2)}_\Omega
=&
\frac{mf^2d^2{\omega'}^2}{8}
\left(
1-\frac{8\pi}{3}\varepsilon^2
\right).
\label{eq:gap2}
\\
\varepsilon_{\text{crit}}^{(2)}=&\sqrt{\frac{3}{8\pi}}\simeq0.35
\label{eq:critcond2}
\end{align}
\end{subequations}

As anticipated at the beginning of Section \ref{sec:secordpert}, the threshold for the onset of coherence is significantly lowered and also the energy gap is more pronounced compared to the calculation at first order.

\section{Spatial dimension of the coherent states and Concept of Natural Resonating Cavity \label{par:SpatDim}}
We have shown the dynamical relevance of quantum states composed of a very large number of elementary charged particles plus a macroscopic (classical) electromagnetic field whose energy content is lower than that of the perturbative state.

The properties of the coherent state \eqref{eq:trial} allow us to identify two vector order parameters $\vec\alpha(t)$ and $\vec{\mathcal A}(t)$ whose temporal evolution is

\begin{subequations}
\begin{align}
\vec \alpha(t)=&
\begin{bmatrix}
\alpha &0&0
\end{bmatrix}
e^{-i \tilde\omega t}
\label{eq:alphat}
\\
\vec{\mathcal A}(t)=&
\begin{bmatrix}
\mathcal A &0&0
\end{bmatrix}
e^{i (\frac\pi2-\tilde\omega t)}
\label{eq:At}
\end{align}
\end{subequations}
where $\alpha\ge0$ and $\mathcal A\ge0$. The choice of the spatial direction 1 is arbitrary.

So far we have neglected the spatial properties of the problem. The reason for this is the very long wavelength of the radiation field when compared to the scale of the lattice. We now want to get some insight into the nature of the spatial properties of the lowest-energy state.
The wave number of the radiation field is unaffected by the coherent transition and is given by 
\begin{equation}
|\vec k|=\omega=\sqrt{{\omega'}^2-\omega_p^2}=\frac{2\pi}{\lambda}
\label{eq:wavenumber}
\end{equation}
 and it is reasonable to expect that the spatial region where a coherent condensation can occur has a minimum radius of the order of the half-wavelength $r_{CD}=\lambda/2=\frac{\pi}{\omega}$. Such a region will be given the name \emph{coherence domain} (CD).

Going back to the definition of the photon field \eqref{eq:radiation}, we can reformulate the expansion in plane waves in terms of spherical harmonics given by \cite{messiah2014quantum}
\begin{equation}
e^{i\vec k\cdot\vec x}=4\pi\sum_{l=0\atop |m|\le l}^\infty 
i^lj_l(|\vec k||\vec x|)Y^*_{lm}(\hat k)Y_{lm}(\hat x).
\end{equation}
By substitution we get
\begin{equation}
\hat{\vec A}(\vec x,t)=
\hat{\vec A}_0(\vec x,t)
+
\delta \hat{\vec A}(\vec x,t)
\label{eq:Asph_harm}
\end{equation}
where
\begin{subequations}
\begin{align}
\hat{\vec A}_0(\vec x,t)=&
\frac{j_0(\omega r)}{\sqrt {2\omega'V}}
\sum_{z=1}^3
(C_z+C_z^\dagger)\hat z
\label{eq:A0}
\\
\delta\hat{\vec A}(\vec x,t)=&
\sum_{l=1\atop |m|\le l}^\infty
\frac{4\pi j_l(\omega r)}{\sqrt {2\omega'V}}
Y_{lm}(\hat x)
\sum_{z=1}^3
\sumkr
[i^lb_{p,\vec k}(t)Y^*_{lm}(\hat k)(\vec\varepsilon_{p,\vec k}\cdot\hat z) \hat z+\text{c.c.}]
\label{eq:A1}
\end{align}
\end{subequations}

\begin{figure} 
\centering
\includegraphics[width=.7\textwidth]{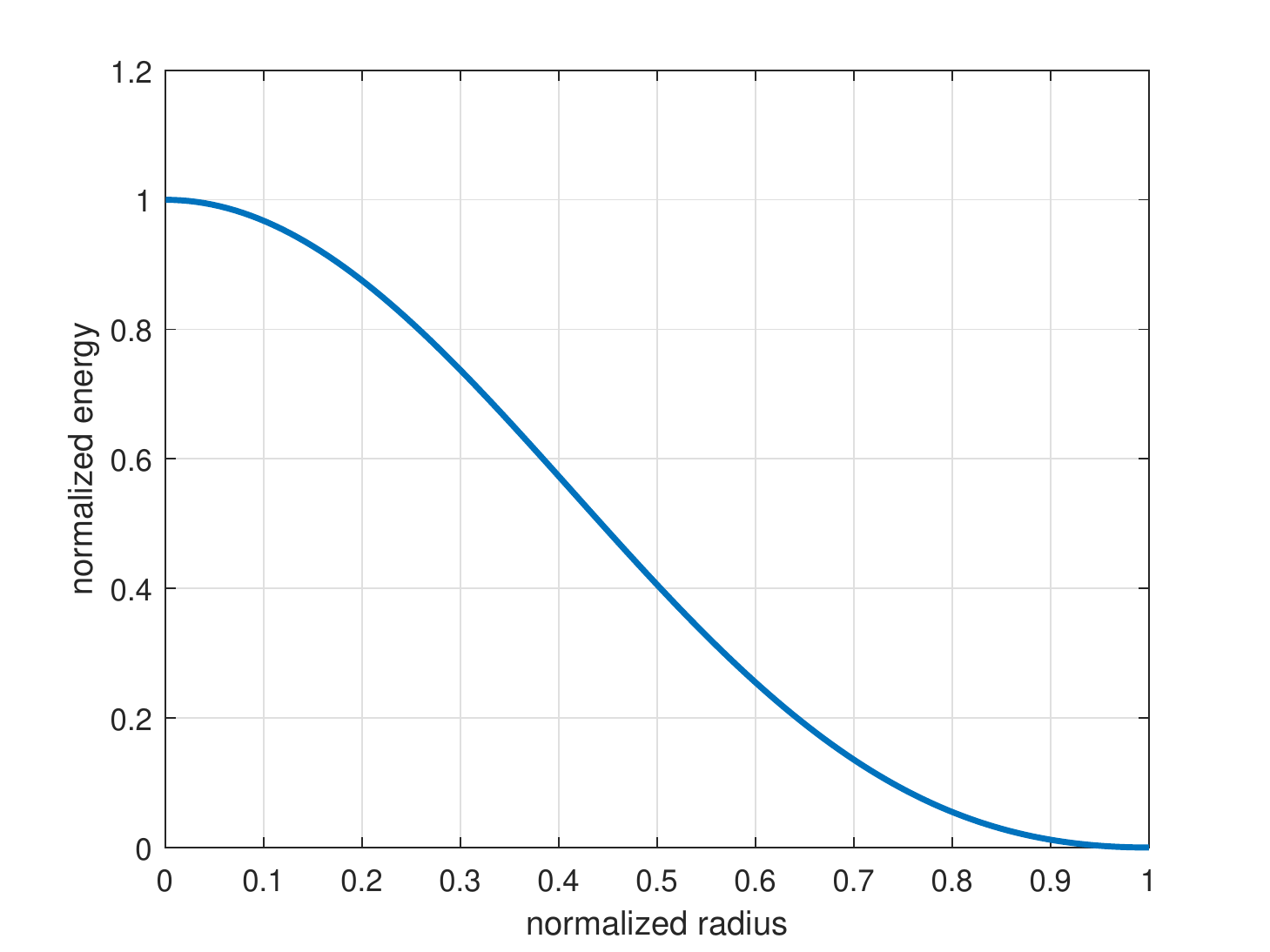}
\caption{Normalized energy profile $j_0^2(\pi\frac{r}{r_{CD}})$ as a function of the normalized radius of the coherence domain.}
\label{fig:energyProfile}
\end{figure}

It can be shown that (see Appendix \ref{app:definition})
\begin{equation}
\bra \Omega \delta\hat{\vec A}(\vec x,t) \ket \Omega
=0,
\end{equation}
therefore 
 the intensity profile of the vector potential is given 
 by the zeroth-order expansion in spherical harmonics of the exponentials in Eq. \eqref{eq:radiation}. Calling $r$ the distance from the center of the CD, we can write
 \begin{equation}
\mathcal A(r)=\mathcal A j_0 \left( \pi\frac r{r_{CD}} \right)
\label{eq:profileA}
\end{equation}
where $j_0(x)$ is the spherical Bessel function of order 0.
The density must be constant throughout the whole CD in order to maintain the frequency (and consequently the quantum phase) spatially constant and being $\alpha$ proportional to $\mathcal A$ (see Eq. \eqref{eq:Avsalpha}), also the profile of $|\alpha(r)|$ is modulated spatially by $ j_0(\pi\frac r{r_{CD}})$. 
If we consider a volume where the charges are distributed uniformly in a sphere of radius $r_{CD}$ with density $\rho=\frac NV=\frac{3N}{4\pi r_{CD}^3}$ the electromagnetic field is zero on the surface of the sphere and varies with the distance from the center following the profile \eqref{eq:profileA}. Thanks to Eq. \eqref{eq:Avsalpha}, the matter amplitude $|\alpha(r)|$ has to be proportional to $|\mathcal A(r)|$  throughout the CD so that the profile of the matter amplitude is given by
 \begin{equation}
\mathcal \alpha(r)=\alpha j_0 \left( \pi\frac r{r_{CD}}\right)
\label{eq:profilealpha}
\end{equation}
and the profile of the energy gap  turns out to be modulated by $ j_0^2(\pi\frac r{r_{CD}})$ (Fig. \ref{fig:energyProfile}). 

Let us now make some considerations that go beyond the exact calculation carried out so far, allowing  the density of the oscillating charges
 and its neutralizing background charge density to vary to some extent (while maintaining its uniformity inside the CD) through migration within the material. We realize that the spatial modulation of the energy gap inside the CD produces a gradient that pulls the charges towards the center of the CD thus increasing the density to the maximum value compatible with the other terms of the energy per particle (that have been neglected in our calculation), namely the electrostatic Coulomb repulsion among particles and Pauli repulsion in case of fermions. As a final result the system reaches an equilibrium density where the attractive coherent potential is counterbalanced by the repulsive short-range terms.

The density on the outside is lower than that on the inside the CD due to the equilibrium of the chemical potential and a sharp variation of the density is present at the boundary of the CD, implying a difference between the plasma frequency inside and outside the CD. The mismatch of the dispersion relations inside and outside the CD implies the total reflection of the EM field at the interface. A natural resonating cavity is thus generated.

In particular, being the dispersion relation of the renormalized radiating field $\omega'=\sqrt{|\vec k|^2+\omega_p^2}$ different from Eq. \eqref{eq:wavenumber}, the electromagnetic field cannot propagate  through the vacuum and remains trapped inside the material. 
 In other words, the described mechanism accounts for the formation of a self-generated, natural QED cavity once the density of the charges overcomes a threshold.

Such configurations may well be identified as new emerging mesoscopic structures able to justify particular properties of matter that display intrinsic quantum features.

The spatial modulation of the fields implies that the average energy gap throughout the CD is given by
 \begin{equation}
     <\delta E>=\delta E(0)\frac{4\pi\int_0^{r_{CD}}r^2dr j_0^2(\pi\frac{r}{r_{CD}})}{\frac43\pi r_{CD}^3}=
     \frac{3}{2\pi^2}\delta E(0)\simeq0.152\ \delta E(0)
 \end{equation}

\section{Discussion}
\label{discussion}

Most of the existing theories of condensed matter do not consider the radiative component of the
electromagnetic field as playing a central role, according to the idea that it is strongly shielded by the high density of the charges  and that only
the electrostatic forces generated by the constituents of matter are important. This is indeed true most of the times, when the attenuation length for the electromagnetic field can be much smaller than the wavelength of the radiating field. However in some circumstances the attenuation  can be compensated by a positive optical gain of the material as it happens, for instance, in solid state lasers and in optical microcavities where exciton-polaritons are formed \cite{byrnes2014exciton,keeling2020bose,mcghee2021polariton}.
A similar mechanism is at play in the problem under study.

In the previous sections we have shown how, at least when temperature is low enough, a spontaneous phase transition is possible for a system of  charges oscillating around equilibrium positions and immersed in a spatially modulated neutralizing charge density of opposite sign, via an instability of the QED vacuum.

In section \ref{par:SpatDim} we have found that the typical spatial dimension of the coherence domains is fixed by the wavelength $\lambda$ of the  radiation field trapped inside the domains. However, this dimension is usually much smaller than the typical dimension of bulk material so that we should expect that the bulk material contains a collection of coherence domains. Such domains are characterized by a single macroscopic wave function, an order parameter and a well-determined quantum phase that confer the domains intrinsically macroscopic quantum properties.

The arguments developed in the present work are valid for bosonic states of matter but remain valid for fermion systems, once a trial state $\ket\Omega$  with the anti-symmetry of the global wave function has been defined. This theoretical development will be addressed in a future work.

Regarding the energetic stability of the solutions found, and besides the fact that in \cite{rokaj2022free} the spatial distribution of matter is two-dimensional while in our case it is three-dimensional, a substantial difference between the results reached in \cite{rokaj2022free} and ours lies in the fact that the states considered in \cite{rokaj2022free} are plane waves, whereas in the present work we have considered coherent states whose spatial size and oscillation remain limited thanks to the intervention of non-linear forces. This fact avoids a divergent energy gap associated with such coherent states.

The physical system analyzed in this work is at zero temperature. The next step will be to consider its thermal properties. This topic will be addressed in a future work but we can anticipate that thermal excitations cause part of the particles forming the coherent state to populate the levels of the single quasi-particle spectrum. The net result is the formation of a two-fluid system similar to that proposed for superfluid helium \cite{tilley2019superfluidity}.
Within this theoretical framework all the properties associated with the degrees of freedom of the quasi-particles of the consolidated condensed matter theory involve the incoherent fraction and only at low temperatures the coherent fraction macroscopically manifests its intrinsically coherent characteristics.

A similar coherent condensation mechanism produced by the electromagnetic field can occur also in atomic or molecular systems having an electric dipole. In this case the electromagnetic field couples to two electronic levels displaying a sufficiently large electric dipole. A very interesting example is represented by liquid water \cite{del1995electrodynamical, w4030510}.

In general, in a physical system made up of different types of charges with a sufficiently high density, various types of electrodynamic coherence associated with different degrees of freedom of the charges can coexist. For instance in a crystal, valence or conduction electrons and ions can form their own coherent states, each one resonating with the electromagnetic field at their specific plasmon polariton frequency. 

\section{Conclusions}
\label{conc}

In this work we have defined and mathematically solved the problem of determining the lowest energy state of a large number of charged bosons coupled to selected modes of the electromagnetic field and harmonically oscillating around their equilibrium positions defined by the vertices of a crystal whose lattice spacing is much smaller than the wavelength of the modes of the electromagnetic field.

The equilibrium positions are determined by a periodic electrostatic potential which we have called the \emph{jellium crystal} which implements global charge neutrality and localization.

The solution found corresponds to a state in which both the electromagnetic and matter fields are coherent and oscillate in quadrature and with a renormalized frequency with respect to the perturbative solution.

The energy content of this state is lower than that of the perturbative state. This is not in contrast with the no-go theorems \cite{andolina2019cavity, nogo2} since our calculation takes into account the contribution of the wave vectors in all directions while the latter consider a single wave vector therefore do not apply to our case.

The calculated energetic gap is not divergent since it depends on the maximum oscillation that the charges can perform which is limited by the finite size of the elementary cells of the crystal.

From a heuristic analysis of the solutions found we define a spatial region called the \emph{coherence domain} (CD) whose size is given by the wavelength of the field and in which both the matter and electromagnetic fields are described by single macroscopic wave functions spatially modulated by the spherical harmonic $j_0(|\vec k|r)$ where $r$ is the distance from the center of the domain and $|\vec k|$ the modulus of the wave vectors of the e.m. field. Due to the modulation of the fields also the energy gap depends on $r$, resulting in an average gap $\braket{\delta E}$ over the coherence domain which is about 15\% of the maximum gap $\delta E(0)$. This average value is numerically evaluated for a case involving protons loaded in the octahedral voids of a typical FCC crystal, giving a gap of $-1$ eV per particle.

The coherent electromagnetic field remains confined within the domain due to the mismatch of the dispersion relation on the edge of the domain and cannot therefore propagate thus forming a natural QED cavity.
If the crystal is spatially larger than the size of a single coherence domain the material will be filled with a collection of CDs.

The present analysis is performed at zero temperature. The thermodynamics of the coherence domains will be presented in a future work.

\appendix

\section{The jellium crystal \label{app:jellium}}
In this Appendix we define a static charge distribution able to reproduce the almost-harmonic potential used in our Hamiltonian.

As was done in the jellium model \cite{hughes2006theoretical}, we assume that the neutralizing charge distribution is static and not perturbed by the presence of the oscillating charges.

In order to simplify the calculation we approximate the cubic cell to a spherical cell with diameter $d$ and assume that in the interstitial regions the charge density is zero. 
In addition, we assume that a uniform negative charge distribution due to the conduction electrons of the metal present in the cell, contributing with a total negative charge of $-e$. We further assume that the oscillating charges contribute with a positive charge $\loadingratio e$ for each cell, where $\loadingratio$ is the hydrogen loading ratio of the metal.

We want to implement charge neutrality and the approximated harmonic potential through a charge distribution $\rho(\xi)$, where $\xi$ is the distance from the center of the cell, such that
\begin{equation}
    4\pi\int^{d/2}_0 \xi^2d\xi \rho(\xi)=-(1+\loadingratio )e
    \label{eq:jel1}
\end{equation}
being $\loadingratio e$ the charge of the single oscillator.

The presence of the oscillators in the centers of the cells represents a localized positive charge that tends to attract the neutralizing charges so as to form an atomic-like distribution of electrons around the equilibrium positions of the oscillators. We model such a distribution with a hydrogen-like plus an uniform charge distribution. The charge density that satisfies these requirements is ($a_0$ is the Bohr radius and $d=2\zeta a_0$ with $\zeta\simeq2.4$ for a typical metal lattice)
\begin{equation}
    \rho(\xi)=-\frac{\loadingratio e}{\pi d^3}
    \left[
    \zeta^3 {\mathrm{e}}^{-2\xi/a_0}+
    {3{\mathrm{e}}^{-\zeta } \,{\left(2\,\zeta +\zeta^2 +2\right)}}
    \right]
    -\frac{6e}{\pi d^3}.
\end{equation}

In the approximation of small oscillation $\xi\ll a_0$ the solution of the Poisson equation in spherical symmetry yields a harmonic potential given by
\begin{equation}
V(\xi)=
\frac{\loadingratio e}{\pi d^3}
    \left[
    \zeta^3/3 +
    {{\mathrm{e}}^{-\zeta } \,{\left(2\,\zeta +\zeta^2 +2\right)}}
    \right]
    \frac{\xi^2}{2} 
    +\frac{2e}{\pi d^3}\frac{\xi^2}{2} 
\label{eq:Poisson1}
\end{equation} 
so that the electrostatic frequency $\omega$ is
\begin{equation}
\omega^2=
\frac{\loadingratio e^2}{\pi md^3}
    \left[
    \zeta^3/3 +
    {{\mathrm{e}}^{-\zeta } \,{\left(2\,\zeta +\zeta^2 +2\right)}}
    \right]
    +\frac{2e^2}{\pi md^3}
\label{eq:Poisson2}
\end{equation} 
and the coupling $g=\frac{\omega_p}{\omega}$ is computed to be
\begin{equation}
g^2=\frac{\pi \loadingratio }{ \loadingratio 
    \left[
    \zeta^3/3 +
    {{\mathrm{e}}^{-\zeta } \,{\left(2\,\zeta +\zeta^2 +2\right)}}
    \right]+2}.
\label{eq:Poisson3}
\end{equation} 
Putting numbers with $\zeta\simeq2.5$ and $x=1$ we find $g\simeq0.61$, $\varepsilon=\frac{g}{\sqrt{1+g^2}}\simeq0.52>\varepsilon_{\text{crit}}$ and
$\omega'=1.92\ \omega_p$.

\section{Renormalization of the photon field \label{app:renorm}}
We define a canonical transformation for the photon creation operators through a diagonalization procedure.
The starting Hamiltonian is
\begin{equation}
\hat H_{\text{photon}}=
\omega
\sumkr
\left[
b_{p,\vec k}^\dagger(t) 
b_{p,\vec k} (t)
+\frac12\right]+
\frac12
\omega_p^2\int d^3\vec x
\hat{\vec A}^2(\vec x,t)
\label{eq:hApp0}
\end{equation}
where we have made the substitution $\sum_{n=1}^N\rightarrow\frac{N}{V}\int d^3\vec x$. In our approximation Eq. \eqref{eq:hApp0} can be written
\begin{equation}
\hat H_{\text{photon}}=
\sumkr
\mathcal H(p,\vec k)
\end{equation}
where
\begin{equation}
    \mathcal H(p,\vec k)=
\omega
\left(
b_{p,\vec k}^\dagger(t) 
b_{p,\vec k} (t)
+\frac12
\right)+
\frac{\omega_p^2}{4\omega}
\left(
b_{p,\vec k}(t)\vec\varepsilon_{p,\vec k} +b_{p,\vec k}^\dagger(t)\vec\varepsilon^*_{p,\vec k} \right)^2.
\label{eq:AppH1}
\end{equation}

By defining the vector operators 
\begin{equation}
\begin{aligned}
B=&\begin{bmatrix}
 b_{p,\vec k}(t) \\   b_{p,\vec k}^\dagger(t)
\end{bmatrix}    
\\
B^\dagger=&\begin{bmatrix}
b_{p,\vec k}^\dagger(t) & b_{p,\vec k}(t)
\end{bmatrix}    
\end{aligned}
\end{equation}

and the matrix
\begin{equation}
    W=
    \begin{bmatrix}
    \frac{\omega}{2}+\frac{\omega_p^2}{4\omega} & \frac{\omega_p^2}{4\omega}
\\
  \frac{\omega_p^2}{4\omega}   & \frac{\omega}{2}+\frac{\omega_p^2}{4\omega}
\end{bmatrix}
\end{equation}
Eq. \eqref{eq:AppH1} is written in a compact form as
\begin{equation}
    \mathcal H(p,\vec k)=
B^\dagger W B.
\label{eq:AppH2}
\end{equation}
By diagonalizing $W$ we find the matrix of the eigenvalues and eigenvectors
\begin{equation}
    U=\frac{1}{\sqrt{2\omega\omega'}}
    \begin{bmatrix}
    -\omega' & \omega \\ \omega' & \omega
    \end{bmatrix}
    \ \ \ \ \ 
    w=\frac{1}{2\omega}
    \begin{bmatrix}
    \omega^2 & 0 \\ 0 & {\omega'}^{2}
    \end{bmatrix}
\end{equation}
where the normalization of the eigenvectors is chosen so that $U^\dagger wU=\frac{\omega'}2\mathbb{I}$ and by defining 
\begin{equation}
\begin{bmatrix}
i\hat p \\ \hat q
\end{bmatrix}=U^{-1}B=\frac{1}{\sqrt{2\omega\omega'}}
\begin{bmatrix}
\omega (b_{p,\vec k}-b_{p,\vec k}^\dagger)
\\
\omega' (b_{p,\vec k}+b_{p,\vec k}^\dagger)
\end{bmatrix}    
\end{equation}
we arrive at the "dressed" destruction and creation operators
\begin{subequations}
\begin{align}
c_{p,\vec k}=&
\frac{1}{\sqrt{2}}
(
\hat q+i\hat p
)
=
\frac{1}{2\sqrt{\omega\omega'}}
[(\omega'+\omega) b_{p,\vec k}
+(\omega'-\omega)b_{p,\vec k}^\dagger
]
\\
c^\dagger_{p,\vec k}=&
\frac{1}{\sqrt{2}}
(
\hat q -i\hat p
)
=
\frac{1}{2\sqrt{\omega\omega'}}
[(\omega'-\omega) b_{p,\vec k}
+(\omega'+\omega)b_{p,\vec k}^\dagger
]
\end{align}
\label{eq:cantrans}
\end{subequations}
and Eq. \eqref{eq:AppH2} can be finally written as
\begin{equation}
    \mathcal H(p,\vec k)=
\omega'\left(c^\dagger_{p,\vec k}c_{p,\vec k}+\frac12\right)
\label{eq:AppH3}
\end{equation}
and Eq. \eqref{eq:hApp0} becomes
\begin{equation}
\hat H_{\text{photon}}=
\omega'
\sumkr
\left(c^\dagger_{p,\vec k}c_{p,\vec k}+\frac12\right).
\end{equation}
The transformations \eqref{eq:cantrans} are canonical and the vector potential can be written in terms of the new operators as
\begin{equation}
\hat{\vec A}(\vec x,t)=
\frac{1}{\sqrt {2\omega'V}}
\sumkr
[c_{p,\vec k}e^{i(\vec k\cdot\vec x-\omega't)}\vec\varepsilon_{p,\vec k}+c_{p,\vec k}^\dagger e^{-i(\vec k\cdot\vec x-\omega't)}\vec\varepsilon^*_{p,\vec k}]
\label{eq:Appradiation}
\end{equation}
with a dispersion relation $\omega'=\sqrt{|\vec k|^2+{\omega_p}^2}$, different from that of the vacuum $|\vec k|=\omega$.

\section{Detailed calculation of the expectation values of the states $\ket{\sigma, \alpha, \mathcal A}_{\vecn}$ 
\label{app:expval}}
For simplicity of notation we set $\mathcal B=\sqrt N\mathcal A$ and ${R}^2=|\alpha|^2+|\mathcal B|^2$ so that
\begin{equation}
H'
\ket{\sigma, \alpha, \mathcal A}_{\vecn}=
i \frac1{2\sqrt N}
\frac{\sqrt{\sigma!}}{{R}^{\sigma}}
\sum_{\eta=0}^\sigma
 \frac{\alpha^{\sigma-\eta}}{\sqrt{(\sigma-\eta)!}}
 \frac{\mathcal B^{\eta}}{\sqrt{\eta!}}
 (a_{\vecn}^\dagger C_1-a_{\vecn} C_1^\dagger)
 \ket{\sigma-\eta}_{\vecn}\otimes\ket\eta
\end{equation}
and after action of the operators
\begin{equation}
\begin{split}
H'
\ket{\sigma, \alpha, \mathcal A}_{\vecn}=&
i \frac1{2\sqrt N}
\frac{\sqrt{\sigma!}}{{R}^{\sigma}}
\sum_{\eta=1}^\sigma
 \frac{ (\sigma-\eta+1)
\alpha^{\sigma-\eta}\mathcal B^{\eta}}{\sqrt{(\sigma-\eta+1)!(\eta-1)!}}
\ket{\sigma-\eta+1}_{\vecn}\otimes\ket{\eta-1}+
\\
-&
i \frac1{2\sqrt N}
\frac{\sqrt{\sigma!}}{{R}^{\sigma}}
\sum_{\eta=0}^{\sigma-1}
 \frac{ (\eta+1)
\alpha^{\sigma-\eta}\mathcal B^{\eta}}{\sqrt{(\sigma-\eta-1)!(\eta+1)!}}
\ket{\sigma-\eta-1}_{\vecn}\otimes\ket{\eta+1}
\end{split}
\end{equation}
By shifting the indexes we get
\begin{equation}
\begin{split}
&H'
\ket{\sigma, \alpha, \mathcal A}_{\vecn}=
\\
=-&
i \frac1{2\sqrt N}
\frac{\sqrt{\sigma!}}{{R}^{\sigma}}
\sum_{\eta=0}^\sigma
 \left[
 (\sigma-\eta)
\alpha^{\sigma-\eta-1}\mathcal B^{\eta+1}
 -\eta
\alpha^{\sigma-\eta+1}\mathcal B^{\eta-1}
\right ]
 \frac{ \ket{\sigma-\eta}_{\vecn}\otimes\ket{\eta}}{\sqrt{(\sigma-\eta)!\eta!}}
\end{split}
\end{equation}
and the expectation value of $H'$ becomes
\begin{equation}
\begin{split}
&_{\vecn}\bra{\sigma, \alpha, \mathcal A}H'
\ket{\sigma, \alpha, \mathcal A}_{\vecn}=
\\
&
i \frac1{2\sqrt N}
\frac{\sigma!}{{R}^{2\sigma}}
\sum_{\eta=0}^\sigma
 \frac{\alpha^{*\sigma-\eta}\mathcal B^{*\eta} 
 \left[
 (\sigma-\eta)
\alpha^{\sigma-\eta-1}\mathcal B^{\eta+1}
 -\eta
\alpha^{\sigma-\eta+1}\mathcal B^{\eta-1}
 \right]}{{(\sigma-\eta)!\eta!}}
=
\\
&
=
\frac{\sigma!}{{R}^{2\sigma}}
\left[
i \frac1{2\sqrt N}
\alpha^{*}\mathcal B 
\sum_{\eta=0}^{\sigma-1}
 \frac{
 (\sigma-\eta)
|\alpha|^{2(\sigma-\eta-1)}|\mathcal B|^{2\eta}
 }{{(\sigma-\eta)!\eta!}}
-
i \frac1{2\sqrt N}
\alpha\mathcal B^{*}
\sum_{\eta=1}^\sigma
\frac{
 \eta
|\alpha|^{2(\sigma-\eta)}|\mathcal B|^{2(\eta-1)}
 }{{(\sigma-\eta)!\eta!}}
 \right]=
\\
&
=
i \frac1{2\sqrt N}
\frac{\sigma!}{{R}^{2\sigma}}
(\alpha^{*}\mathcal B 
-\alpha\mathcal B^{*})
\sum_{\eta=1}^\sigma
\frac{
|\alpha|^{2(\sigma-\eta)}|\mathcal B|^{2(\eta-1)}
 }{{(\sigma-\eta)!(\eta-1)!}}=
 \\
 &
=
- \frac1{\sqrt N}
\frac{\sigma!}{{R}^{2\sigma}}
|\alpha\mathcal B |
\sum_{\eta=1}^\sigma
\frac{
|\alpha|^{2(\sigma-\eta)}|\mathcal B|^{2(\eta-1)}
 }{{(\sigma-\eta)!(\eta-1)!}}
 =
-
|\alpha\mathcal A |
\frac{\sigma}{{R}^2}
\end{split}
\end{equation}
The expectation values of the matter Hamiltonian in the $\vecn$ sector is
\begin{equation}
{_{\vecn}\bra{\sigma, \alpha, \mathcal A}}
H
\ket{\sigma, \alpha, \mathcal A}_{\vecn}
=\frac{\sigma!}{{R}^{2\sigma}}
\sum_{\eta=0}^\sigma
 \frac{
 |\alpha|^{2(\sigma-\eta)}|\mathcal B|^{2\eta}
 \left[
 \sigma-\eta+\frac32
 \right]}{{(\sigma-\eta)!\eta!}}
=
\omega'
\left[
\sigma
 \frac{|\alpha|^2}{R^2}+\frac32
 \right]
\end{equation}
and of the photon Hamiltonian

\begin{equation}
\begin{split}
&_{\vecn}\bra{\sigma, \alpha, \mathcal A}
\hat H_{\text{photon}}
\ket{\sigma, \alpha, \mathcal A}_{\vecn}=
\\
&
\omega'
\frac{\sigma!}{{R}^{2\sigma}}
\sum_{\eta=0}^\sigma
 \frac{
 |\alpha|^{2(\sigma-\eta)}|\mathcal B|^{2\eta}
 \left[
(\eta+\frac32)
 \right]}{{(\sigma-\eta)!\eta!}}
=
\\
&=
\omega'
\left[
\frac{\sigma!}{{R}^{2\sigma}}
|\mathcal B|^{2}
\sum_{\eta=1}^\sigma
 \frac{
 |\alpha|^{2(\sigma-\eta)}|\mathcal B|^{2(\eta-1)}
}
 {{(\sigma-\eta)!(\eta-1)!}}+\frac32
 \right]=
\\
&=
\omega'
\left[
\frac{\sigma!}{{R}^{2\sigma}}
|\mathcal B|^{2}
\sum_{\eta=0}^{\sigma-1}
 \frac{
 |\alpha|^{2(\sigma-1-\eta)}|\mathcal B|^{2\eta}
}
 {{(\sigma-1-\eta)!\eta!}}+\frac32
 \right]=
\\
&=
\omega'
\left[
\frac{\sigma}{{R}^{2}}
N|\mathcal A|^2
+\frac32
 \right]
\end{split}
\end{equation}

\section{An estimate of the numerical value of the energy gap
\label{estimate}}

Let us make an estimate of the energy gap per particle for an ensemble of protons adsorbed into a metallic matrix. The fermionic nature of protons can be neglected since the Fermi energy at the density we are considering is negligible. 
Assume that the lattice spacing is $d=2.5\angstrom=2.5\cdot 5.0674\cdot10^{-4}$ eV$^{-1}=0.0013$ eV$^{-1}$ and that there is one proton for each cell.
The mass of a proton is $m\simeq938$ MeV and the coupling with the electromagnetic field is $\varepsilon=0.52$, as has been computed in Appendix \ref{app:jellium}. The system is above threshold, therefore coherent.
Finally we choose in (\ref{eq:firstorder}) and (\ref{eq:gap2}) $f\simeq0.4$, which is the typical relative dimension of the octahedral voids in a metallic lattice.
Inserting the numerical values we obtain
\begin{equation}
\begin{split}
\omega_p=&\sqrt{\frac{e^2}{md^3}}=0.22 \text{  eV}
\\
\omega'=&\ 0.41 \text{  eV}
\\
\omega=&\ 0.35 \text{  eV}
\\
\Delta E(0)=&-6.8 \text{  eV}
\\
<\Delta E>=&-1.0 \text{  eV}
\end{split}
\end{equation}
The computed energy gap is in the range of the chemical energies and is a good candidate for the description of the spontaneous hydrogen absorption observed in many metals. Moreover, once the analysis is extended to electrons, where the full fermionic nature must be taken into account, this mechanism could have relevance in the theoretical description of the work function. 

\section{Calculation of $\bra \Omega \delta\hat{\vec A}(\vec x,t) \ket \Omega$
\label{app:definition}}
We start from the definition \eqref{eq:radiation} and we expand the exponentials in spherical harmonics so that
\begin{equation}
\hat{\vec A}\ (\vec x,t)=
\sum_{l=0}^\infty
\hat{\vec A}_l\ (\vec x,t)
\end{equation} 
where we have defined
\begin{subequations}
\begin{align}
\hat{\vec A}_l\ (\vec x,t)=&
j_l(\omega r)Y_{lm}(\hat x)
\sum_p
\int d\Omega_{\hat k}
[\vec f_l(p,\hat k)b_{p,\vec k}(t)+\text{c.c.}]
\\
\vec f_l(p,\hat k)
=&
\frac{4\pi}{\sqrt {2\omega'V}}
\sum_{|m|\le l}
i^lY^*_{lm}(\hat k)\vec\varepsilon_{p,\vec k}.
\end{align}
\end{subequations}
We now compute the expectation value of $\hat{\vec A}_l\ (\vec x,t)$ on the coherent state $\ket{\mathcal A}$. We need to evaluate the expectation value of the $n-$th term of the sum of Eq. \eqref{eq:defB}
\begin{equation}
\sum_{p, p_1,... p_n=1}^2
\int 
d\Omega_{\hat k}
d\Omega_{\hat k_1}...
d\Omega_{\hat k_n}
f_l^*(p,\hat k)f_0(p_1,\hat k_1)... f_0(p_n,\hat k_n)
\bra{\mathcal A}
b^\dagger_{p,\vec k}
b^\dagger_{p_1,\vec k_1}
...
b^\dagger_{p_n,\vec k_n}
\ket0.
\end{equation}
After performing the contractions, the only term of the sum that survives is the $n+1-$th and is proportional to $\int d\Omega_{\hat k} f_l^*(p,\hat k)f_0(p,\hat k)$ that yields zero for $l>0$, thanks to the orthogonality of the spherical harmonic functions.

Likewise, following the same argument, we find for $l>0$
\begin{equation}
\sum_{p, p_1,... p_n=1}^2
\int 
d\Omega_{\hat k}
d\Omega_{\hat k_1}...
d\Omega_{\hat k_n}
f_l(p,\hat k)f_0(p_1,\hat k_1)... f_0(p_n,\hat k_n)
\bra{\mathcal A}
b_{p,\vec k}
b^\dagger_{p_1,\vec k_1}
...
b^\dagger_{p_n,\vec k_n}
\ket0=0
\end{equation}
and we get finally
\begin{equation}
\bra{\mathcal A}
\hat{\vec A}_l(\vec x,t)
\ket{\mathcal A}=0 \ \ \ \text{for } l>0.
\end{equation}

\bigskip
\bigskip
\noindent
\textbf{Author Contribution Statement -} All authors contributed equally to this paper.

\bigskip
\bigskip
\noindent
\textbf{Data Availability Statement -} Data sharing is not applicable to this article as no datasets were generated or analysed during the current study.

\bibliography{coherent_plasma} 

\begin{thebibliography}{10}

\bibitem{rokaj2018light}
V~Rokaj, DM~Welakuh, M~Ruggenthaler, and A~Rubio.
\newblock Light--matter interaction in the long-wavelength limit: no
  ground-state without dipole self-energy.
\newblock {\em Journal of Physics B: Atomic, Molecular and Optical Physics},
  51(3):034005, 2018.

\bibitem{andolina2019cavity}
GM~Andolina, FMD Pellegrino, V~Giovannetti, AH~MacDonald, and M~Polini.
\newblock Cavity quantum electrodynamics of strongly correlated electron
  systems: A no-go theorem for photon condensation.
\newblock {\em Physical Review B}, 100(12):121109, 2019.

\bibitem{andolina2020theory}
GM~Andolina, FMD Pellegrino, V~Giovannetti, AH~MacDonald, and M~Polini.
\newblock Theory of photon condensation in a spatially varying electromagnetic
  field.
\newblock {\em Physical Review B}, 102(12):125137, 2020.

\bibitem{ashida2020quantum}
Y~Ashida, A~{\.I}mamo{\u{g}}lu, J~Faist, D~Jaksch, A~Cavalleri, and E~Demler.
\newblock Quantum electrodynamic control of matter: Cavity-enhanced
  ferroelectric phase transition.
\newblock {\em Physical Review X}, 10(4):041027, 2020.

\bibitem{guerci2020superradiant}
D~Guerci, P~Simon, and C~Mora.
\newblock Superradiant phase transition in electronic systems and emergent
  topological phases.
\newblock {\em Physical Review Letters}, 125(25):257604, 2020.

\bibitem{stokes2020uniqueness}
A~Stokes and A~Nazir.
\newblock Uniqueness of the phase transition in many-dipole cavity quantum
  electrodynamical systems.
\newblock {\em Physical Review Letters}, 125(14):143603, 2020.

\bibitem{mivehvar2021cavity}
F~Mivehvar, F~Piazza, T~Donner, and H~Ritsch.
\newblock {Cavity QED with quantum gases: new paradigms in many-body physics}.
\newblock {\em Advances in Physics}, 70(1):1--153, 2021.

\bibitem{roman2021photon}
J~Rom{\'a}n-Roche, F~Luis, and D~Zueco.
\newblock {Photon condensation and enhanced magnetism in cavity QED}.
\newblock {\em Physical Review Letters}, 127(16):167201, 2021.

\bibitem{rokaj2022free}
V~Rokaj, M~Ruggenthaler, FG~Eich, and A~Rubio.
\newblock Free electron gas in cavity quantum electrodynamics.
\newblock {\em Physical Review Research}, 4(1):013012, 2022.

\bibitem{nogo2}
P~Nataf and C~Ciuti.
\newblock {No-go theorem for superradiant quantum phase transitions in cavity
  QED and counter-example in circuit QED}.
\newblock {\em Nat. Commun.}, 1(72):510--532, 2010.

\bibitem{lauscher2000rotation}
O~Lauscher, M~Reuter, and C~Wetterich.
\newblock Rotation symmetry breaking condensate in a scalar theory.
\newblock {\em Physical Review D}, 62(12):125021, 2000.

\bibitem{branchina1999antiferromagnetic}
V~Branchina, H~Mohrbach, and J~Polonyi.
\newblock {Antiferromagnetic $\varphi$4 model. I. The mean-field solution}.
\newblock {\em Physical Review D}, 60(4):045006, 1999.

\bibitem{modanese1998stability}
G~Modanese.
\newblock {Stability issues in Euclidean quantum gravity}.
\newblock {\em Physical Review D}, 59(2):024004, 1998.

\bibitem{bonanno2013modulated}
A~Bonanno and M~Reuter.
\newblock Modulated ground state of gravity theories with stabilized conformal
  factor.
\newblock {\em Physical Review D}, 87(8):084019, 2013.

\bibitem{bonanno2019structure}
A~Bonanno.
\newblock On the structure of the vacuum in quantum gravity: A view from the
  asymptotic safety scenario.
\newblock {\em Universe}, 5(8):182, 2019.

\bibitem{modanese2021quantum}
G~Modanese.
\newblock {Quantum metrics with very low action in R+R2 gravity}.
\newblock {\em Physical Review D}, 103(10):106020, 2021.

\bibitem{burns1993mineralogical}
RG~Burns.
\newblock {\em Mineralogical applications of crystal field theory}.
\newblock Number~5. Cambridge University Press, 1993.

\bibitem{faisal1987theory}
FHM Faisal.
\newblock {\em Theory of multiphoton processes}.
\newblock Springer Science \& Business Media, 1987.

\bibitem{wang2004thermodynamic}
Y~Wang, Z-K Liu, and L-Q Chen.
\newblock {Thermodynamic properties of Al, Ni, NiAl, and Ni3Al from
  first-principles calculations}.
\newblock {\em Acta Materialia}, 52(9):2665--2671, 2004.

\bibitem{ref5}
G.~Preparata.
\newblock {\em QED coherence in matter}.
\newblock World Scientific, 1995.

\bibitem{frisk2019ultrastrong}
Anton Frisk~Kockum, Adam Miranowicz, Simone De~Liberato, Salvatore Savasta, and
  Franco Nori.
\newblock Ultrastrong coupling between light and matter.
\newblock {\em Nature Reviews Physics}, 1(1):19--40, 2019.

\bibitem{BrillouinWigner}
PM~Morse and H~Feshbach.
\newblock {\em Methods of Theoretical Physics, Vol. II}.
\newblock McGraw-Hill Book Company Inc., 1953.
\newblock Page 999.

\bibitem{messiah2014quantum}
A~Messiah.
\newblock {\em Quantum mechanics}.
\newblock Courier Corporation, 2014.

\bibitem{byrnes2014exciton}
T~Byrnes, NY~Kim, and Y~Yamamoto.
\newblock Exciton--polariton condensates.
\newblock {\em Nature Physics}, 10(11):803--813, 2014.

\bibitem{keeling2020bose}
J~Keeling and S~K{\'e}na-Cohen.
\newblock {Bose--Einstein condensation of exciton-polaritons in organic
  microcavities}.
\newblock {\em Annual Review of Physical Chemistry}, 71:435--459, 2020.

\bibitem{mcghee2021polariton}
KE~McGhee, A~Putintsev, R~Jayaprakash, K~Georgiou, ME~O’Kane, RC~Kilbride,
  EJ~Cassella, M~Cavazzini, DA~Sannikov, PG~Lagoudakis, et~al.
\newblock Polariton condensation in an organic microcavity utilising a hybrid
  metal-dbr mirror.
\newblock {\em Scientific Reports}, 11(1):1--12, 2021.

\bibitem{tilley2019superfluidity}
DR~Tilley and J~Tilley.
\newblock {\em Superfluidity and superconductivity}.
\newblock Routledge, 2019.

\bibitem{del1995electrodynamical}
E~del Giudice, A~Galimberti, L~Gamberale, and G~Preparata.
\newblock Electrodynamical coherence in water: A possible origin of the
  tetrahedral coordination.
\newblock {\em Modern Physics Letters B}, 9(15):953--961, 1995.

\bibitem{w4030510}
I~Bono, E~Del~Giudice, L~Gamberale, and M~Henry.
\newblock Emergence of the coherent structure of liquid water.
\newblock {\em Water}, 4(3):510--532, 2012.

\bibitem{hughes2006theoretical}
RIG Hughes.
\newblock {Theoretical practice: the Bohm-Pines quartet}.
\newblock {\em Perspectives on Science}, 14(4):457--524, 2006.

\end{thebibliography}
\bibliographystyle{unsrt}

\end{document}